\documentstyle[preprint,aps,psfig] {revtex}
\begin{document}

\title {Pionic Fusion of Heavy Ions}

\author{
D. Horn, 
G.C. Ball,
D.R. Bowman, 
W.G. Davies,
D. Fox,\\
A. Galindo-Uribarri,
A.C. Hayes,
and
G. Savard}
\address{AECL, Chalk River Laboratories, Chalk River, Ontario, Canada
K0J 1J0}

\author{
L. Beaulieu,
Y. Larochelle,
and
C. St-Pierre}
\address{Laboratoire de physique nucl\'eaire, D\'epartement de physique, 
Universit\'e Laval,\\ Ste-Foy, Qu\'ebec, Canada G1K 7P4}

\date{\today}
\maketitle

\begin{abstract}
We report the first experimental observation of the pionic fusion of two heavy
ions. The {$^{12}$C}({$^{12}$C},{$^{24}$Mg}){$\pi^0$}\ and
{$^{12}$C}({$^{12}$C},{$^{24}$Na}){$\pi^+$}\ cross sections have
been measured to be $208\pm 38$ and $182\pm 84$ picobarns, respectively, at
$E_{cm} = 137$ MeV. This cross section for heavy-ion pion production, at an
energy just 6 MeV above the absolute energy-conservation limit, constrains
possible production mechanisms to incorporate the kinetic energy of the entire
projectile-target system as well as the binding energy gained in fusion.
\end{abstract}
\vspace{0.5in}
PACS number(s): 25.70.-z, 25.60.Pj, 25.75.Dw, 13.60.Le

\newpage

Pionic fusion, the creation of a pion in a reaction fusing two nuclei, requires
the concentration of the kinetic and potential (binding) energy of the
projectile-target system into a few degrees of freedom, namely the rest mass
and emission energy of a pion, with possible, discrete emission energies
determined by the limited number of available final states of the fusion 
product. The phenomenon has previously been observed for the
{$^3$He}({$^3$He},{$\pi^+$}){$^6$Li}\
reaction\cite{leb81} and subsequently for {$^3$He}-induced reactions with
heavier targets \cite{bim84,sch86}. Until now, it has been a matter of
conjecture whether reactions between two heavy ions would manifest the degree
of coherence necessary for pionic fusion. By contrast, the inclusive creation
of pions in heavy-ion reactions  at ``subthreshold'' energy ({\em i.e.} at an
energy per nucleon below the nucleon-nucleon pion production threshold) has
been studied extensively (see refs. \cite{bra87} and \cite{cas90} for reviews).
However, low experimental efficiencies for pion detection make measurements of
sub-nanobarn cross sections difficult; as a result experiments have been
limited to center-of-mass energies 100 MeV or more above absolute threshold
\cite{you86,wat93}. Even at those energies, mechanisms based upon on-shell
nucleon-nucleon collisions\cite{cas90} could not
account for the observed pion production cross sections, and a
number of cooperative or coherent mechanisms were proposed, including ``pionic
bremsstrahlung'' \cite{vas84} and statistical compound nucleus
emission of pions\cite{pra86,pot88}. To date, the application of fully
quantum-mechanical models for pion production in fusion reactions has
been limited to cases in which the projectile is a proton\cite{fea82}.
Motivated by the successful descriptions\cite{gop74} of the
$p+p \rightarrow d + ${$\pi^+$}\ reaction, these models usually involve three
basic ingredients: a primary nucleon-nucleon pion production
mechanism, an amplitude for rescattering this virtual pion on-shell, and a
nuclear structure form factor describing the required spectroscopic amplitudes
and the momentum distributions of the wave functions. An extension of these
microscopic models to the present
{$^{12}$C}+{$^{12}$C}\ reaction, which has projectile/target symmetry
and an isoscalar entrance channel, necessarily requires at
least two nucleons in the pion production amplitude.
Then, in the simplest picture, one of the {$^{12}$C}\ nuclei may be excited to a
T=1 state producing a virtual pion, and the pion then rescattered on-shell from
the second {$^{12}$C}. The pseudoscalar nature of the pion production vertex
means that only unnatural-parity T=1 states of {$^{12}$C}\ are involved. Thus
the cross section depends sensitively on the nuclear structure spectroscopic
amplitudes for expressing the final states of the mass-24 product in terms of
the coupling of T=1 one-particle-one-hole unnatural parity states of {$^{12}$C}\
to the {$^{12}$C}\ ground state.

This Letter reports the first observation of heavy-ion pionic fusion. The very
low center-of-mass energy relative to threshold requires a coherent production
process. The cross sections may be linked with the existing systematics of
inclusive {$^{12}$C}({$^{12}$C},{$\pi^0$}) measurements at higher
energies\cite{bra87,cas90}.
The pionic fusion channel, in particular, provides a reference point for
calculations of various proposed statistical or thermal production processes,
since the initial conditions (compound nucleus mass, charge, and excitation
energy) are determined by the fusion process. In addition, the present
restriction to an isoscalar entrance channel provides a constraint on existing
microscopic models, which may lead to an understanding of the dominant reaction
mechanism at ``subthreshold'' energies.

In order to measure cross sections in the sub-nanobarn range, where
{$\pi^0$}-decay studies are hampered by background
and efficiency considerations, and charged-pion
detection is limited by spectrometer solid
angle, we have chosen to detect the mass-24
fusion products of the {$^{12}$C}({$^{12}$C},{$^{24}$Mg}){$\pi^0$}\
and {$^{12}$C}({$^{12}$C},{$^{24}$Na}){$\pi^+$}\
reactions, rather than the pions themselves.
Any {$^{24}$Mg}\ compound nucleus formed without
emitting a high-energy quantum, such as a pion or a bremsstrahlung photon,
should decay by nucleon or cluster emission, giving a lower mass. High-energy
photon emission can be excluded kinematically by the large momentum it
transfers to the recoil. This makes mass-24 recoils, moving with the
approximate center of-mass velocity of the reaction, a reliable signature of
pionic fusion. At $E_{cm}=137$ MeV, the total energy available is only about
6 MeV above that necessary to produce a positive or neutral pion and the
corresponding $A=24$ isobar in its lowest $T=1$ state; the {$\pi^-$}\ channel
is energetically forbidden.

A {$^{12}$C}\ beam of $E_{lab}=274.2\pm 1.2$ MeV from Chalk River's TASCC
facility was incident on a thick, isotopically separated {$^{12}$C}\ target,
826 $\mu$g/cm$^2$ in areal density. The target was prepared with precautions to
minimize contamination, baked in a vacuum oven, and stored and transferred in
an inert atmosphere. With the assumption of a maximum-energy 6-MeV pion emitted
transversely and with allowance for multiple scattering in the target, the
{$^{24}$Na}\ and {$^{24}$Mg}\ products recoil within one degree of the beam
axis. We therefore aligned the Chalk River Q3D spectrometer \cite{mil70} at
$\theta_{lab}=0${$^\circ$}\ and collimated its entrance with a circular
aperture of 1 msr solid angle, corresponding to a cone of 1{$^\circ$}\
half-angle. The main carbon beam was stopped and current-integrated on a beam
block positioned inside the spectrometer's magnet box; additional cleanup of
scattered beam was provided by a scraper paddle further into the spectrometer.
Ions with the appropriate momentum-to-charge ratio were deflected
to the spectrometer's focal plane and registered in a heavy-ion counter
\cite{bal79}. The first layer of the detector was an avalanche counter
operating at 10 Torr and providing timing and position information. The
avalanche counter was built in two halves, with a dead region between them
reducing the total detector efficiency to 94\%. This layer was followed by a
second gas volume at about 200 Torr, which contained proportional wires for
determination of energy loss and position, an ionization chamber volume to
measure residual energy, and finally an anticoincidence
region to reject events arising from ions (mainly scattered or degraded beam
particles), failing to stop in the $E$ detector. Events electronically
imitating mass-24 recoil signals, such as several ions reaching the counter at
nearly the same time, were largely eliminated by timing, pileup rejection, and
anticoincidence requirements. A more severe problem was presented by genuine
{$^{24}$Na}\ and {$^{24}$Mg}\ ions formed in reactions with
heavier-than-carbon target contaminants (chiefly oxygen); their presence
necessitated a detailed examination of background yields. The momentum
calibration was obtained from the position signals produced with a 138-MeV
{$^{24}$Mg} \ beam for various spectrometer magnetic fields. The
charge-state distributions for 120- to 150-MeV {$^{24}$Mg}\ ions exiting a
carbon target were explicitly measured (more than half were fully stripped)
and the results extrapolated for sodium. Since
the momentum acceptance of the focal plane detector spanned about 4.7\%, four
spectrometer field settings were needed to cover the momentum range in the
vicinity of the beam momentum for mass-24 ions of charge 11 and 12, and data
from 7500 $\mu$C of {$^{12}$C}$^{6+}$ beam ions were collected at each setting.  

The galilean-invariant yields for {$^{24}$Na}\ and {$^{24}$Mg}\ are shown as a
function of velocity in the upper and lower panels, respectively, of Fig. 1.
The spectra are not background-subtracted but are corrected for dead-time
and detector efficiency. The expected widths of the velocity
distributions, the calibration uncertainties, resolution effects, 
energy-loss straggling in the target, and the
possible range of beam energies combine to determine
the ``allowed'' velocity ranges for pionic fusion residues (indicated by the
double-ended arrows), over which the yield is integrated
to obtain the gross cross sections listed in the first column of Table 1.

Note the non-zero yields outside of the ``allowed'' velocity ranges, especially
the rise in yield toward velocities well below the center-of-mass velocity of
the {$^{12}$C}-{$^{12}$C}\ system. Such an increase at low velocity would be
characteristic of evaporation residues from fusion reactions with target
contaminants heavier than carbon. Analysis of the target composition by the ERD
technique \cite{for95} revealed the presence of oxygen contamination at the 1\%
level and a smaller amount of aluminum contamination. To obtain the spectral 
shape of the background due to oxygen, the major contaminant, within the
allowed velocity region, we have measured the
{$^{16}$O}({$^{12}$C},{$^{24}$Na}) and {$^{16}$O}({$^{12}$C},{$^{24}$Mg}) 
yields from a 150-$\mu$g/cm$^2$ MoO$_2$ target. We then obtained the
{$^{24}$Na}\ and {$^{24}$Mg}\ background levels by normalizing the $^{23}$Na
and $^{23}$Mg yields from the oxygen reaction to those
from the carbon target, where the main source of mass-23 ions is contamination.
(Single-nucleon emission from the {$^{12}$C}+{$^{12}$C}\ reaction would cause
the A=23 residue to recoil outside the region of velocity space allowed for
pionic fusion residues.)
The resulting background levels are indicated by shading in Fig. 1.
Their shape is that of the oxygen contaminant products, but their yield, from
the normalization to mass-23 carbon-target products, represents the {\em total}
contaminant product cross section for {$^{24}$Na} and {$^{24}$Mg}. The background
cross sections, integrated over the same velocity range as the 
carbon-target data, are listed in the table and used in the determination of the
net yields for the pionic fusion process. The background cross sections
obtained are consistent with those interpolated from the mass-24 yield on
each side of the ``allowed'' region.

With decreasing velocities, the carbon-target data apparently
rise even more steeply than the oxygen-target background measurement.
Heavier contaminants, such as aluminum, produce mass-24 residues more readily
than oxygen does, and would give such a low-velocity rise,
but these products have little impact in the velocity
range of pionic fusion residues, since their distributions are
peaked at even lower velocities than those from oxygen contamination.
We have made a third independent background
assessment based on the shape of the {$^{25}$Mg} spectrum, which samples all
the heavier-than-carbon contaminants on our carbon target. When normalized to 
the mass-24 yields at $v_{recoil}=0.103c$, below the allowed velocity range
for pionic fusion residues, the backgrounds obtained by this method (dotted 
lines in the figure) agree with the {$^{24}$Na} and {$^{24}$Mg} 
oxygen background measurements listed in the table to within 30 picobarns.
The consistency between the three independent background evaluations
demonstrates that the contaminant effects are quantitatively understood
and confirms the reliability of the background-subtracted cross sections.

The shape of the recoil velocity distribution carries information about the
kinematics of the emitted pion. The well-known forward-backward peaking
observed in {$^{12}$C}({$^{12}$C},{$\pi^0$}) reactions at higher energies
\cite{gro85} should combine with the phase-space preference for higher
emission energies to produce a depletion in recoil yield at the
center-of-mass velocity. The primary distribution of recoil velocities
resulting from full-energy pion emission ({\em i. e.} to the lowest T=1
state) and a cosine-squared pion angular distribution as measured in the
{$^3$He}({$^3$He},{$\pi^+$}){$^6$Li}\ reaction\cite{leb81} is plotted in 
the top panel of Fig. 2. The lineshape expected after energy loss for 
{$^{24}$Mg} recoils produced over an 826-$\mu$g/cm$^2$  range of target
thickness and smeared by straggling, resolution effects, and final-state
gamma-ray emission is drawn as a heavy dashed line; for this target
thickness the depleted velocity region is filled in. The lineshape for
a thinner target (solid line) retains its central valley. The appropriate
thick-target lineshapes, added to the measured background spectra, and
postioned arbitrarily within the allowed velocity range, are superimposed
on the data of Fig. 1. 

To search experimentally for the kinematic signature
of the pionic fusion process, additional
measurements were made with thinner, 486-$\mu$g/cm$^2$ targets.
Their sum is shown in the lower half of Fig. 2, illustrating
the characteristic depletion in {$^{24}$Mg}\ yield at the center-of-mass
velocity. The cross section for the thin-target measurements are  listed in
Table 1. The {$^{24}$Mg}\ results are in agreement with
thick-target data, but the {$^{24}$Na}\ yields, for which only one
low-statistics experiment is available, are inconclusive.
Since the maximum of the low-velocity peak
falls on the dead space joining two halves of the timing detector for that
reaction, an additional 35\% of systematic error has been included in the
tabulated gross value. For such a brief run, the net cross section expected
on the basis of the thick-target data (227 pb) should have a total uncertainty
of $\pm 198$ pb, indicating that our actual experimental result
of $23\pm 90$ pb has little statistical significance. The result does,
however, lower the weighted average cross section by some 20\% relative
to the thick-target data alone, and should therefore not be ignored in
the averaging.

The radiative capture reaction (fusion followed by emission of gamma-rays
only) could also contribute {$^{24}$Mg}\ ions at these velocities.
It was therefore decided to use the same target to measure the
{$^{24}$Mg}\ yield at $E_{cm} = 130$ MeV, which is just below the absolute
threshold for the {$^{12}$C}({$^{12}$C},{$^{24}$Mg}[T=1]){$\pi^0$}\
reaction. At this energy, the radiative capture cross section should be
essentially unchanged, but pionic fusion is forbidden. The
background-subtracted cross section is $59 \pm \ 109$ pb, which is consistent 
with zero (within half a standard deviation) and inconsistent (by 1.4 standard
deviations) with the average above-threshold cross section. With this level
of experimental uncertainty, the radiative capture process cannot be 
conclusively eliminated on the basis of our data; however, a number
of additional factors argue against it, such as the shape of the thin-target
velocity distribution, the kinematic exclusion of single-photon emission, and 
the relative improbability of multiple high-energy photon emission (statistical-model calculations\cite{cha95} are three orders
of magnitude lower than the measured cross sections). Furthermore, radiative
capture is not a possible source of $^{24}$Na ions.

To assess the significance of the cross sections in terms of overall heavy-ion
pion production, it is necessary to know what other pionic exit channels are
open. For the {$\pi^+$}\ channel, the only final states allowed by energy
conservation involve particle-stable levels in {$^{24}$Na}. The
{$^{12}$C}({$^{12}$C},{$^{24}$Na}){$\pi^+$}\ yield of $182 \pm \ 84$ pb,
averaged from the thick-target and thin-target measurements, therefore
represents the entire {$\pi^+$}\ cross section. (Because of the disparity
between the measurements, we follow the common practice of inflating the
uncertainty on the mean by a factor of $\sqrt{\chi ^2/N_F}$.)
The {$\pi^0$}\ channel has a larger energy range for allowed final states
in {$^{24}$Mg}, but the upper half of this range is particle-unbound.
A more detailed accounting of the specific T=1 levels available, their
particle decay widths, and phase space factors indicates that about half the
{$^{24}$Mg}\ residues produced in the {$\pi^0$}\ reactions survive. This,
together with the expected isospin relationship for the {$\pi^+$}\ and
{$\pi^0$}\ channels and the effect of the Coulomb barrier on {$\pi^+$}\ decay,
should result in a mass-24 residue yield that is comparable for the two pion
channels. The average measured {$^{12}$C}({$^{12}$C},{$^{24}$Mg})
cross section of $208 \pm \ 38$ pb is therefore consistent with this estimate.
Extrapolation of the inclusive {$\pi^0$}\ production
systematics\cite{bra87,cas90} to lower energy should be reliable to within
a factor of 2 or 3 and  gives 500 pb for our reaction. The measured cross
section, representing about half the inclusive yield, is in agreement with
these systematics. The impact of the experimental results on some of the
currently available models is illustrated below:
\begin{itemize}
\item Standard heavy-ion particle-production models ({\em e.g.} ref. 
\cite{cas90}), based on incoherent summation of on-shell nucleon-nucleon
collisions, are appropriate at higher energies, but severely underpredict our
near-threshold results.
\item The pionic bremsstrahlung model \cite{vas84} requires \ $E_{cm}\geq
m_{\pi}c^2+E_{Coul}$, \ since the relatively gentle deceleration from the
Coulomb barrier (about 10 MeV for {$^{12}$C} +{$^{12}$C} ) should not
contribute to pion production. With $E_{cm}\approx m_{\pi}c^2$, the present
pionic fusion reactions would not have sufficient kinetic energy to produce
pionic bremsstrahlung after Coulomb deceleration.
\item A test of various thermal/statistical models \cite{pra86,pot88} should
also be possible with pionic fusion data. For inclusive pion production, the
thermal approach depends on models of reaction dynamics to define the size and
excitation of the emitting volume; for pionic fusion, these quantities are
known. Furthermore, with a two-body exit channel, the pion emission energy is
defined by the known low-lying $T=1$ states in the final nucleus.
\item Microscopic models\cite{fea82,gib82} require the $T=0$ target or
projectile to be excited to a $T=1$ particle-hole state; experimental
observation of pions in the present reaction necessitate at least two
nucleons in the pion production amplitude. In addition, pionic fusion of two
heavy ions pushes the models further off shell than do proton-induced
reactions.
\end{itemize}

In summary, the {$^{12}$C}({$^{12}$C},{$^{24}$Mg}){$\pi^0$}\ and
{$^{12}$C}({$^{12}$C},{$^{24}$Na}){$\pi^+$}\ cross sections
have been measured to be $208\pm 38$ and 182$\pm 84$ picobarns, respectively,
at $E_{cm} = 137$ MeV. This constitutes the first observation of the pionic
fusion of two heavy ions. It also provides a low-energy measurement, just
6 MeV above absolute threshold, for the extensive heavy-ion {$\pi^0$}\
production systematics, which had only been studied previously at energies
more than 100 MeV above threshold. Our measurement of appreciable yields so
near threshold will require models to incorporate coherent mechanisms for
pion production.

We thank Mark Chadwick and Bill Gibbs for valuable discussions. This work has
been partially funded by the Natural Sciences and Research Council of Canada
and by AECL.

\begin{table}
\vspace{1in}
\caption{Gross, background, and background-subtracted cross sections for the 
{$^{12}$C}({$^{12}$C},{$^{24}$Mg}) and {$^{12}$C}({$^{12}$C},{$^{24}$Na})
reactions with thick (826-$\mu$g/cm$^2$)
and thin (486-$\mu$g/cm$^2$) targets at $E_{cm} = 137$ MeV in the region
of ``allowed'' recoil velocities for pionic fusion. Also listed is the
subthreshold ($E_{cm} = 130$ MeV) thin-target measurement for the
{$^{12}$C}({$^{12}$C},{$^{24}$Mg}) reaction.
}
\end{table}

\begin{figure}
\vspace{1in}
\caption{Galilean-invariant cross sections as a function of recoil velocity,
obtained from the thick-target data, for the reactions
{$^{12}$C}({$^{12}$C},{$^{24}$Na}) (top panel) and
{$^{12}$C}({$^{12}$C},{$^{24}$Mg}) (bottom panel) at
$E_{cm} = 137$ MeV. For consistency, data are binned by position on the
spectrometer focal plane, rather than by analyzed velocity. 
The double-headed arrows indicate the possible range
of recoil velocity distributions from the {$\pi^+$}\ (top) and {$\pi^0$}\
(bottom) pionic fusion reactions. The shaded regions represent the measured
A=24 yields from the {$^{12}$C}+{$^{16}$O}\
reaction, scaled to represent the background due to oxygen contamination
of the target; the dotted line is an alternative background determination based
on {$^{25}$Mg}\ yields. The solid curves represent one possible calculated
lineshape, illustrating the expected width. See text for details and lineshape
calculation.
}
\end{figure}

\begin{figure}
\vspace{1in}
\caption{Top panel: Velocity lineshape expected for pionic fusion residues
assuming a cos$^2\theta$ pion angular distribution (dotted line). The
broadening from recoil energy loss in the target, straggling,
resolution effects, and final-state gamma decay are shown for target
thicknesses of 486 $\mu$g/cm$^2$ (solid line) and 826 $\mu$g/cm$^2$ (dashed 
line). See text for description of lineshape calculation. Bottom panel:
As Fig. 1, but for {$^{12}$C}({$^{12}$C},{$^{24}$Mg}) thin-target data.
}
\end{figure}

\newpage

\hspace*{0.in}
\begin{minipage}{5.in}
\begin{table}
\vspace*{1.5in}
\begin{tabular}{|c|c|c|c|c|} 
\ Measurement \
& counts \  
&\ {\Large $\sigma_{gross}$}(pb) \
&\ {\Large$\sigma_{bkgd}$}(pb) \
&\ {\Large$\sigma_{net}$}(pb) \ \\
\hline
\ {$^{24}$Mg}\ (thick)
&\ 98 
&\ 397(40)
&\ 166(25)
&\ 231(47)\\
\hline
\ {$^{24}$Na}\ (thick)
&\ 68
&\ 353(43)
&\ 126(22)
&\ 227(48)\\
\hline
\ {$^{24}$Mg}\ (thin)
&\ 58
&\ 329(43)
&\ 163(46)
&\ 166(63)\\
\hline
\ {$^{24}$Na}\ (thin)
&\ 7
&\ 159(82)
&\ 136(38)
&\ 23(90)\\
\hline
{$^{24}$Mg}\ (subthresh)
&\ 14
&\ 279(90)
&\ 220(62)
&\ 59(109)\\
\end{tabular}
\vspace{3in}
\center{Table\ I.\ D. Horn {\em et al.}}
\end{table}
\end{minipage}

\begin{figure}
\vspace*{1in}
\psfig{figure=pilet1.epsf,width=6.5in,height=5.5in}
\vspace*{1in}
\center{Fig.\ 1.\ D. Horn {\em et al.}}
\end{figure}

\begin{figure}
\vspace*{1in}
\psfig{figure=pilet2.epsf,width=6.5in,height=5.5in}
\vspace*{1in}
\center{Fig.\ 2.\ D. Horn {\em et al.}}
\end{figure}

\end{document}